\documentclass[10pt,a4paper]{article} 
\usepackage[dvips]{graphicx}
\usepackage{amsfonts,amsmath,amssymb}
\usepackage{hyperref}
\usepackage{cite}

\newcommand{\bse}{\begin{subequations}}
\newcommand{\ese}{\end{subequations}}

\begin{document}
\title{\bf Rotating Dense Quark Matter and Anomaly Inflow }

\date{}
\maketitle
\vspace*{-0.3cm}
\begin{center}

{\bf  Majid Dehghani$^{}$}\\

{\it {Research Institute for Astronomy and Astrophysics of Maragha (RIAAM), \\
P.O. Box 55134-441, Maragha, Iran
}} \\

\vspace*{0.3cm}
{\it   {${}$m.dehghani55@gmail.com} }
\end{center}

\begin{abstract}

 The effective theory of rotating pion superfluid in the presence of topological defects will be considered.
 We study the anomaly induced effects and the interplay between domain-wall and superfluid vortex under rotation. A non-uniform rotation leads to new effects in the domain-wall and vortex system. It will be shown that the effective theory predicts radial current flows of charges in the system whereas the previously studied cases dealt with induced static charges. The main observation is that the radial flow consists of two parts which are related to the presence of gauge and gravitational anomalies. The microscopic picture of fermionic zero modes propagating along the vortices will be used to justify the chiral effective theory results and clarifies the mechanism for the current flow. Then gravitoelectromagnetic formalism is used to redrive the gauge anomaly related part of radial flow. Finally, as our main observation,  we discuss the $(1+1)$-dimensional gravitational anomaly on the vortex ring which entails an energy-momentum flow on the domain-wall. Again the results will be confirmed from a microscopic point of view.

\end{abstract}

\section{Introduction}

  Recently rotating systems of quark matter under extreme condition have attracted much attention.
  The motivation for current interest lies in recent reports on alignment of $\Lambda$ hyperon polarization and global angular momentum of noncentral heavy ion collision \cite{STAR:2017ckg}.
  The topological current induced by rotation of matter is called chiral vortical effect (CVE). It was originally concerned in  astrophysical context for Dirac matter in rotating frame \cite{Vilenkin1980}. The CVE effect is considered to be the macroscopic manifestation of the triangle anomaly in the underlying chiral theory. It is closely related to chiral magnetic effect (CME) i.e. current generation along an external magnetic field induced by chirality imbalance. In fact interesting analogy between rotation and magnetic field exists\cite{Kharzeev:2007tn}. The CME and CVE are investigated in different frameworks ranging from hydrodynamics to microscopic zeromode calculations on topological defects. Here we will consider the CVE as it may be relevant for the description of the dense quark matter under extreme conditions in rotating neutron stars, and noncentral heavy ion collisions with rotating fireballs.

   The CVE was identified in chiral kinetic theory \cite{Stephanov:2012ki}, qualitatively in holographic models  \cite{Erdmenger:2008rm, Banerjee:2008th, Torabian:2009qk} and then in \cite{Son:2009tf} obtained by considering the entropy current, using the hydrodynamic approximation alone.
  Hydrodynamics has been successful in describing vast variety of phenomena. There are attempts to drive the hydrodynamic theory results in the presence of anomalies for chiral superfluids using effective field theories and anomaly matching. One example is combined action of the chiral perturbation theory and the Wess-Zumino-Witten (WZW) term to study the anomalous effects in chiral superfluids \cite{Lublinsky:2009wr, Lin:2011aa},  where the former is the low energy effective theory of QCD vacuum and the latter corresponding to the chiral anomaly.
  Dense quark matter with high baryon density under rotation is one of the main possible cases for investigating CVE. It is known that many global symmetries of QCD are spontaneously broken at large baryon densities \cite{Alford:2001dt}. The spontaneous breaking of global symmetries leads to the existence of topological defects such as domain walls and strings \cite{Son:2000fh,Forbes:2001gj}.
  Domain-walls are interpolating configurations between two vacua. It is a configuration in which the local expectation value of Goldstone boson varies over a scale of its Compton wavelength along the direction of magnetic field or axis of rotation.  It is supposed that the high-density domain-walls exist and the relevant physics is under theoretical control \cite{Son:2007ny}. Another topological object present in various superfluid phases is the vortex. Because the velocity field of the superfluid is known to be a potential flow, vorticity can be non-vanishing only on a discrete set of lines. So vortices are one dimensional quantum objects. Since vortices carry angular momentum, a rotating superfluid consists of lattice of vortices.
  Closed loop of vortex string can exist and is referred to as a vorton. A superconducting or superfluid string has a persistent current and the resulting finite angular momentum stabilises the vorton by angular momentum conservation \cite{Buckley:2002mx}.

  Many unusual effects was predicted in \cite{Son:2004tq} to take place in the dense QCD matter due to anomalies. Using an effective Lagrangian for the interaction of Goldstone bosons with the magnetic field at finite chemical potential, it was shown that electric charge and magnetization will be induced on axial vortices and domain walls, respectively.
  Axial domain walls exist at very high densities where instanton effects are suppressed \cite{Son:2000fh}. The response of QCD ground state at finite baryon density to a strong magnetic field was considered in \cite{Son:2004tq, Son:2007ny}, where the triangle anomaly causes the coupling of neutral Goldstone boson and magnetic field. As a result, there are anomaly induced magnetization and baryon number densities on the axial domain wall surface. The magnetic field and the resulting induced magnetization are perpendicular to the domain-wall and at sufficiently high magnetic fields the domain wall becomes locally and globally stable \cite{Son:2007ny}.
   It was argued in \cite{Son:2004tq} that the above mentioned phenomenon can be understood in a different way by considering the axial domain wall to be bounded by an axial vortex, where anomaly induced charges on the domain-wall can be related to the charges of vortex ring.
   The relation uses the fact that an axial vortex loop carries angular momentum by placing it inside a rotating quark matter.
  The configuration is called a drum vorton, which was first found in a nonlinear sigma model at finite temperature \cite{Carter:2002te} and then for dense matter in a magnetic field \cite{Son:2004tq}.
  On the other hand, one can perform  microscopic argument for currents on magnetic flux tubes at finite chemical potential which consists of propagating fermionic zero modes on the tube.  The anomaly induced currents for domain-walls bounded by vortices under magnetic field and interplay between them as a manifestation of Callan-Harvey mechanism was investigated in \cite{Fukushima:2018ohd}. It was shown that the presence of vortex singularities requires new terms that otherwise gauge invariance and conservation law would be violated. Also it was pointed out that for a changing magnetic field, axial vortices absorb or emit the charge onto domain-wall. This method is closely related to the previously known approach which relies on the connection between field theory topological defects and anomalies, resulting in the well-known Callan-Harvey anomaly inflow \cite{Goldstone:1981kk, Nielsen:1983rb, Callan:1984sa}.

  Many of the above considerations for domain walls and vortices in a magnetic field can be applied to dense quark matter under rotation.
  Similar to the case of magnetic field, the same reasoning applies for existence of a domain wall surrounded by a vortex in rotating quark matter, i.e. the minimal surface area surrounded by the superfluid vortex ring should form a domain-wall to minimize the energy cost by topological winding.
  The stability and layered structure along with anomaly induced angular momentum, baryon and isospin charges was considered in \cite{Huang:2017pqe} for pion domain walls in rotating dense matter . Superfluidity at finite chemical potential is used as it deals with  important issues such as confinement and chiral symmetry breaking. Superfluidity at finite chemical potential plays a central role in the following discussion and it has been discussed  for different phases of QCD and QCD-like theories.
  QCD-like theories at finite baryon chemical potential and the resulting baryon superfluid was discussed in \cite{Kogut:1999iv, Kogut:2000ek}.
  The pion superfluidity arises at finite isospin chemical potential and zero temperature \cite{Son:2000xc} and a study of simultaneous presence of baryon and isospin chemical potentials was performed in \cite{Metlitski:2005db}. The pion superfluidity induced by isospin chemical potential and the propagating fermionic zero modes along the vortices was considered in \cite{Kirilin:2012mw}.
  Analogy between the fluid rotation and electromagnetic fields was first emphasized in \cite{Kharzeev:2007tn}.

  Now we continue to present our results for non-uniform rotation with new current flows.
  The system of pion domain-wall surrounded by a superfluid vortex and their anomalous behavior under a non-uniform rotation is the main focus of this work.
  We will rely on hydrodynamic approximation to describe rotating superfluid and its topological defects such as vortices.
  Pionic medium under uniform rotation was studied in \cite{Huang:2017pqe}, where the prominent phenomena was induced charges on the domain-wall.
  But the prominent phenomena for non-uniform rotation, as will be explained, is the radial current flow on the domain-wall towards the edge.
  Our main objective is that the radial flow on domain-wall due to non-uniform rotation, consists of gauge and gravitational anomaly related parts.
  The first case elaborated here is the radial flow due to the gauge anomaly in domain-wall and vortex system.
  Two different viewpoints will be used to study gauge anomaly for non-uniform rotation.
  First the local velocity of fluid is treated as an effective gauge field, as introduced in \cite{Sadofyev:2010is}, see also \cite{Abramchuk:2018jhd, Kirilin:2012mw}. The anomalies of theory are obtained via triangle anomaly common to all systems put in a background $U(1)$ gauge field.
  The second method to study the gauge anomaly part is to use the curved metric of a rotating frame \cite{Huang:2017pqe}. In this method the anomalies of effective theory are obtained via anomaly matching. The microscopic theory gives a complementary picture for effective theory results. The effective theory results and propagating fermionic zero mode calculations are in agreement. Our method also gives a microscopic  explanation for the previously known cross-correlation between rotation and magnetic field.

  The next new effect in the rotating system with the specified topology, is the radial flow due to gravitational anomaly.  Pure gravitational anomalies occur in $4k+2$ space-time dimensions, which reflects a failure of reparametrization invariance \cite{AlvarezGaume:1983ig}. The reparametrization non-invariance means nonzero covariant derivative for energy-momentum tensor, $\nabla_{\mu}T^{\mu\nu} \neq 0$.
   It was argued in \cite{Stone:2012ud} that a thermal Hall current  on the domain-wall is possible only, for a non-uniform gravitational field.
  Here, we show that a non-uniform gravitational field is generated by a non-uniform rotation which then results in the failure of conservation law for energy-momentum tensor. This new effect leads to an energy-momentum current flow towards the boundary. We give a microscopic picture of pure gravitational anomaly based on energy-momentum two point function and non-invariance of the fermion effective action under infinitesimal general coordinate transformation. The microscopic argument makes it clear why the pure gravitational anomaly exists only for non-uniform rotation. Then as a manifestation of Callan-Harvey anomaly inflow, an energy-momentum current flows towards the boundary ring. It will be shown that the $(1+1)$-dimensional gravitational anomaly on the vortex ring and the higher dimensional domain-wall Chern-Simons term, leads to energy-momentum current flow.
    It is worth noting that there could be current flows for uniform rotation,  such as the anomalous Hall current  predicted in \cite{Huang:2017pqe}, due to the chemical potential gradient. But the anomalous Hall current is a peripheral rather that radial flow. It exists also for the domain-wall and vortex configuration, where the chemical potential gradient is a consequence of finiteness of domain-wall size (see section \textbf{\ref{rotat frame}}).

   An analogy can be made between our results for non-uniform rotation and the CVE of free fermion systems. In the later, the current flow in the direction of rotation (to linear order) consists of two parts, $j = (T^2/12 + \mu^2/4\pi^2)\Omega$, where the chemical potential and the temperature dependent coefficients are related to gauge and mixed gauge-gravitational anomalies, respectively. Also in our case of concern, the radial flow can be divided into gauge and pure gravitational anomaly related parts. The similarity is only schematic, because the radial flow on the domain-wall and vortex system is perpendicular to the axis of rotation with a different mechanism.
   Here, we mention a constraint for rigid rotation of a system due to causality. In a system rotating with angular velocity $\Omega$, the velocity of a point located at the distance $r$ from the axis of rotation $v=r \Omega$, should not exceed the speed of light, so $r \Omega \leq 1$.

  This paper is organized as follows: In Sec. \textbf{\ref{anom effective}}  we will focus on the effective theory for pion superfluid. The local velocity field will be treated as an effective gauge field and anomalies arising from triangle anomalies. We discuss the implications of rotation for a domain-wall bounded by a superfluid vortex. The main result is the existence of a radial energy  flow due to non-uniform rotation. In Sec. \textbf{\ref{micro argu}} we will analyse the implications from microscopic point of view. First the mechanism for generation of fermion modes in non-uniform rotation will be discussed and then propagating fermionic zero modes on the vortex.
  Sec. \textbf{\ref{rotat frame}} is devoted to studying gravitoelectromagnetic formalism which confirm the previous result and shows that the flow is a Hall type radial flow. We also comment on the peripheral Hall energy flow at the edge of finite size domain-wall.
  Pure gravitational anomaly of two-dimensional vortex ring resulting from a non-uniform rotation and consequent anomaly inflow from higher dimension is the subject of  Sec. \textbf{\ref{grav anom}}.

\section{Anomalies in effective theory}\label{anom effective}

 The two-flavor chiral perturbation theory (ChPT) is characterized by triplet of pseudoscalar Nambu-Goldstone bosons.
 It is concerned to study the dynamical effect of finite isospin chemical potential which leads to pion superfluidity \cite{Son:2000fh, Teryaev:2017}.
 In chiral perturbation theory the effective Lagrangian for pion medium with nonzero isospin chemical potential is of the form

\begin{equation}\label{chiral lagrang}
  \mathcal{L}= \frac{1}{4} f_{\pi}^2 Tr [D^{\mu}U (D_{\mu}U)^{\dag}],
\end{equation}
 where $D_{\mu}U= \partial_{\mu}U - i\mu_I \delta_{\mu0}(\tau_3U -U\tau_3)/2  $, $U$ is unitary matrix of pseudoscalar Nambu-Goldston boson

\begin{equation}\label{unitart pion}
  U= exp \left(\frac{i \tau^a \pi^a}{f_{\pi}} \right).
\end{equation}
 The potential energy corresponding to Lagrangian (\ref{chiral lagrang}) is given by

\begin{equation}\label{effective potential}
  V_{eff}(U)= \frac{f_{\pi}^2 \mu_I^2}{8} Tr[\tau_3U\tau_3U^{\dag}-1],
\end{equation}
 so the vacuum of theory instead of $U=1$ is described by  $U=e^{i\tau_3\phi}$. The degeneracy of (\ref{effective potential}) on $\phi$ signals the
 presence of a Goldstone boson. A fixed value of angle $\phi$ corresponds to spontaneous symmetry breaking and it can be identified with the Goldston boson. Also the well-known time dependence of superfluid phase is

\begin{equation}\label{superfluid phase}
  U_{vac}= e^{i\tau_3(\mu_I t + \varphi(x))}.
\end{equation}
 It is worth noting that in the chiral limit, one can introduce also a chiral chemical potential $\mu_I^5$. As pointed out in \cite{Teryaev:2017}, cases with $(\mu_I\neq 0,\mu_I^5=0)$ and $(\mu_I= 0,\mu_I^5\neq 0)$ can be reduced to each other.
 To study vortices we rely on the hydrodynamic approximation. The hydrodynamics of a charged superfluid incorporating the Goldstone field is worked out in \cite{Son:arxiv2002}.
 We follow \cite{Son:2004tq} to obtain effective theory anomalous terms. The method is to put the system in background gauge fields and require the effective theory to reproduce the anomalies of the microscopic theory. The quark chemical potential is represented as the zeroth components of a fictitious vector gauge field $V_{\mu}$. In the presence of electromagnetic gauge field $A_{\mu}$ and fictitious field  $V_{\mu}$ the conservation of chiral current is violated by triangle anomalies \cite{Son:2004tq, Kirilin:2012mw}  
  
 \begin{equation}\label{AV tringle anomaly}
  \partial^{\mu} j_{\mu} = -\frac{1}{16 \pi^2} \left( e^2 C_{A\gamma \gamma} F^{\mu\nu} \tilde{F}_{\alpha\beta} - 2e C_{AV \gamma} V^{\mu\nu} \tilde{F}_{\alpha\beta} +     C_{A VV}   V^{\mu\nu} \tilde{V}_{\alpha\beta} \right),
\end{equation}
 where $F_{\mu\nu}=\partial_{\mu}A_{\nu}-\partial_{\nu}A_{\mu}$, $V_{\mu\nu}=\partial_{\mu}V_{\nu}-\partial_{\nu}V_{\mu}$, and and $j_{\mu}$ is the axial current which creates the Nambu-Goldstone (NG) boson. Here we have only rotation, so the first and second terms are absent (first term related to the well-known NG boson-two photon coupling and the second term which is a mixed term).
  Then the third term gives the coupling of quarks to the local velocity field $V_{\mu}$, and takes the following form

\begin{equation}\label{tringle anomaly}
  \partial^{\mu} j_{\mu} = -\frac{1}{4 \pi^2} C_{\pi VV} \epsilon^{\mu\nu\alpha\beta}  V_{\mu\nu} V_{\alpha\beta},
\end{equation}
  where $V_{\mu}= \mu v_{\mu}$ with $v_{\mu}=(1,\vec{v})$ . The coefficient $C_{\pi VV}$ depends on current quantum numbers and for simplicity one flavor case is considered here with
 $C_{\pi VV}=1$ \cite{Kirilin:2012mw}. The two flavor effective field theory, where the calculation of coefficient $C_{\pi VV}$ contains trace over Pauli matrices was discussed in \cite{Sadofyev:2010is}.
 Now let us comment on the approximations made in obtaining the effective action (\ref{tringle anomaly}). As explained in the introduction an expansion in velocity and chemical potential for the effective theory was performed in \cite{Sadofyev:2010is}. The anomalous terms are exhausted by the triangle graph or by terms quadratic in chemical potential. Because the anomaly is one loop exact, higher order terms in gradient expansion doesn't alter the gauge anomaly current. The situation is different in a curved background. In hydrodynamics, derivative expansions are considered in weakly curved backgrounds in which the metric expansion is performed around the flat space-time metric.
 Curved metric perturbations come into play at second order of hydrodynamic derivative expansion \cite{Baier:2007ix, Bhattacharyya:2008ji}. So curvature corrections are expected to appear at one loop order for chiral anomaly in a curved background.  It was shown in \cite{Flachi:2017vlp} that the chiral vertical effect in a curved background receives a curvature correction proportional to Riemann curvature tensor. The results are obtained by considering the rotation as a small perturbation in the curved background with the simplifying assumption that all metric components be time independent. The simplifying assumptions of \cite{Flachi:2017vlp} are not valid her but one can expect that more complicated calculations result in curvature corrections to the chiral anomaly.

  Due to the spontaneous breaking of isospin symmetry the low-energy dynamics contains a NG boson $\varphi$, which is the $U(1)$ phase of the condensate. The effective low-energy description must respect the $U(1)$ isospin gauge symmetry. Therefore, to have a covariant derivative in the effective Lagrangian, the derivative $\partial_{\mu}\varphi$ should always appear in conjunction with $V_{\mu}$ \cite{Son:2004tq}.
 From the condition of anomaly matching one can deduce that the effective Lagrangian takes the following form

\begin{equation}\label{anomaly lagrang}
  L_{anom}= \frac{1}{16\pi^2}   \epsilon^{\mu\nu\alpha\beta} ~ \partial_{\mu}\varphi ~ V_{\nu} V_{\alpha\beta}.
\end{equation}
 so for the effective action we have

\begin{equation}\label{4dim anom action}
  S_{anom}= -\frac{1}{16\pi^2} \int d^4x \epsilon^{\mu\nu\alpha\beta} ~ \partial_{\mu}\varphi ~ V_{\nu} V_{\alpha\beta} = -\int d^4x V_{\mu}j_{bulk}^{\mu},
\end{equation}
 where the axial-vector current $j_{bulk}^{\mu}$ is

\begin{equation}\label{current from action}
  j_{bulk}^{\mu}= \frac{\delta S_{anom}}{\delta  V_{\mu}}=\frac{1}{16\pi^2} \epsilon^{\mu\nu\alpha\beta} ~  \partial_{\nu}\varphi ~ V_{\alpha\beta},
\end{equation}
    where $v_{\mu}$ is the local velocity and  vorticity of the form $\partial_i v_j-\partial_j v_i= 2\epsilon_{ijk}\Omega_k $. Then the anomalous action (\ref{4dim anom action}) assumes  the following form

\begin{equation}\label{fluid anomaly}
  S_{anom}= \frac{\mu^2}{16\pi^2} \int d^4x ~~  \boldsymbol{\Omega} . \boldsymbol{\nabla}\varphi.
\end{equation}
  This term accounts for anomaly induced angular momentum and isospin density on the domain-wall.
  Another interpretation will be given later in terms of vortex zero modes.
  If there is no singularity in the superfluid velocity field, the current defined in (\ref{current from action}) is conserved $(\partial_{\mu}j_{bulk}^{\mu}=0)$. But in the presence of a vortex flux, which has a singularity in the superfluid velocity field, the current is no longer conserved

\begin{equation}\label{current nonconserv}
  \partial_{\mu}j_{bulk}^{\mu} \neq 0.
\end{equation}
 A vortex flux singularity leads to nonvanishing $[\partial_{\mu},\partial_{\nu}]\varphi$, which otherwise would be zero. As explained in the introduction in a rotating medium there would be stable pion domain-wall surrounded by vortex ring. The vortex ring along the disc shaped domain-wall causes the above mentioned current nonconservation and $[\partial_i,\partial_j ]\varphi= 2\pi \delta^{(2)}(x_{\bot})$. Taking divergence of (\ref{current from action}), we have

\begin{equation}\label{4D current div}
  \partial_{\mu}j_{bulk}^{\mu}=\frac{1}{16\pi^2} \epsilon^{\mu\nu\alpha\beta} V_{\mu\nu} \partial_{\alpha} \partial_{\beta}\varphi.
\end{equation}
 Next, we  explain that the bulk current non-conservation is cured by vortex zero mode contribution. The induced anomalous charge can be obtained from equation (\ref{current from action})
\begin{equation}\label{bulk zero current}
 j_{bulk}^{0}= \frac{\mu}{8\pi^2} \vec{\Omega}.\vec{\partial}\varphi.
\end{equation}
 and the spatial volume integration gives

\begin{equation}\label{bulk charge}
 Q_{bulk}= \int d^3x ~ j_{bulk}^{0}= \frac{\mu}{4\pi} \int dx^i~v^i = \frac{\mu}{4\pi} \oint_{vortex} d^2x ~ \Omega^3.
\end{equation}
 The zero mode contribution will be obtained in the next section and takes the following form

\begin{equation}\label{zeromode charge}
  Q_{zm}= \int d^3x ~ j_{zm}^{0}= - \frac{\mu}{4\pi} \oint_{vortex} d^2x ~\Omega^3.
\end{equation}
 which is the quantum number for vorticity and cancels the bulk contribution (\ref{bulk charge}). So the conserved total charge is the sum of bulk and zeromode charges

\begin{equation}\label{total charge}
  Q= Q_{bulk}+ Q_{zm}.
\end{equation}
 It is more interesting to study the anomalous current associated with Eq.(\ref{current from action}), which takes the form

\begin{equation}\label{radial current}
  j_{bulk}^r = \frac{1}{4\pi^2} \epsilon^{rt\theta z} V_{t\theta}~\partial_z \varphi.
\end{equation}
 The current exists for nonzero peripheral acceleration, $V_{t\theta}=\mu \partial_t v_{\theta} \neq 0$, and  this happens for a non-uniform rotation.
 So a non-uniform rotation results in a radial current flow on the domain-wall. This conclusion will be confirmed using gravitoelectromagnetic formalism.
 Now as in the case of superconducting string discussed in \cite{Witten:1984eb}, we argue that the vortex is a superfluid ring. For non-uniform rotation there would be a linear acceleration resulting from a peripheral force field $F_{\theta}$, $( V_{01}=\mu\partial_t v_{\theta} \propto F_{\theta}\neq0 )$. The force field exists during the finite period of non-uniform rotation and vanishes for uniform rotation. Although the force field exist for a limited period of time, the current remains. The conservation of the current can be deduced from topological nature of vorticity quantum number (\ref{zeromode charge}), and guarantees persistence of the current. In the next section the effect of induced force field on fermi energies of left- and right-handed zero modes will be explained.

\section{Microscopic Argument} \label{micro argu}

 In this section we give the microscopic picture based on propagating zero modes, following \cite{Kirilin:2012mw, Metlitski:2005pr}. Topology of the system is circular ring geometry discussed in \cite{Fukushima:2018ohd}.
 As explained in the introduction, the fermion zero modes on vortex string can be used to give another interpretation of the phenomena occur in domain-wall and vortex system. We will show that the contributions from effective coupling of zero modes to the velocity field on the vortex and the bulk current non-conservation, cancel each other. Then using microscopic picture of vortices and Hamiltonian index theorem, the vortex quantum number and fermion zero mode changes will be related to each other.
 Previous arguments for relating vortex quantum number and fermion zero modes will be generalized to the case of non-uniform rotation. Unlike uniform rotation, the Hamiltonian for a non-uniform rotation is not constant of motion. Then to be able to apply the formalism of a uniformly rotating system, we can assume a step by step increase in angular momentum. In this section, first we explain the microscopic mechanism for generation of fermionic modes on the vortex and then continue to relate these changes via the index theorem, to the vortex topological winding number changes.
 First Following \cite{Kharzeev:2007tn}, we explain the microscopic mechanism for fermionic mode generation in a non-uniform rotation. The important point is that an exact relation exists for the generation of electric field along the rotation axis in the presence anomalous effective Lagrangian \cite{Son:2004tq}. The result can be generalized for the system with an axion field  \cite{Metlitski:2005pr}. It has been shown that in the presence of an axion field or anomalous effective Lagrangian, an electric field is generated in presence of an external magnetic field or rotation.   For our case of concern, i.e. an anomalous effective Lagrangian with nonzero chemical potential in the presence of rotation, an electric field will be induced along the axis. Anomaly generates disparity and the resulting disparity generates a real electric field in presence of rotation which results in a current along the vortex ring. As the electric field is generated, the mechanism is in accordance with the superconducting string discussed in \cite{Witten:1984eb}, where the induced electric field shifts the Fermi energies of left- and right movers in opposite direction and results in an increase of chiral fermions directed along the electric field.
 We can repeat the same argument for the superfluid system. There are equal number of right- and left-mover fermions in the ground state of superfluid vortex ring. As explained above in the presence of rotation and triangle anomaly, an electric field would be generated which increases the number of, for example, right-movers by elevating the right-movers Fermi energy and decreasing the oppositely moving fermions. This results in creating fermion modes with particular chirality and an increase in total angular momentum.

 As pointed out in \cite{Son:2004tq}, the anomalous domain-wall magnetization also can be interpreted as the magnetic field generated by the current running along vortex loop. In an analogous way the domain-wall angular momentum density is interpreted as the angular momentum of rotating fermions along the closed vortex loop.
 The action of the system with coupling to local velocity field is of the following form

\begin{equation}\label{action vlocity coupl}
  S=  \int d^4x ~\bar{\psi} (\partial_{\mu}+ iV_{\mu})\gamma^{\mu} \psi.
\end{equation}
For simplicity we assume that the rotation velocity remains much smaller than unity so that the radial dependence of contraction factor can be neglected $\gamma(r) \approx 1$  \cite{Abramchuk:2018jhd}, so the chemical potential in the laboratory reference frame $\mu_{lab}=\gamma(r) \mu \approx \mu$.
Now, the zeromode two-dimensional effective action with an effective coupling to the effective gauge field, becomes

\begin{equation}\label{zeromode action}
  S_{zm}= \int dt d\vartheta ~~ \bar{\xi}[i\gamma^a(\partial_a + i\mu v_a )]\xi,
\end{equation}
where $a=1,2$ is the longitudinal string coordinate, $\vartheta=r\theta$. Here $\xi$ is the solution for free Dirac equation and $(\partial_0 \pm \partial_{\vartheta})\xi =0$ with plus sign for right-handed and minus sign for left-handed fermion.
The (1+1)-dimensional gauge anomaly is known to be of the form

\begin{equation}\label{2D gauge anomaly}
  \partial_{\alpha}j^{\alpha5}=-\frac{e}{2\pi}\epsilon^{\alpha\beta}F_{\alpha\beta}=-\frac{e}{2\pi}E
\end{equation}
As the local velocity is treated as an effective gauge field ($eA_{\mu} \rightarrow \mu v_{\mu}$), then the (1+1)-dimensional theory has current non-conservation of the form

\begin{align}\label{covar anom curr}
  \partial_{\alpha}j_{zm}^{\alpha} &=  -\frac{1}{2\pi} V_{01} \delta^{(2)}(x_{\perp}) \nonumber  \\
  &= -\frac{1}{8\pi} \epsilon^{\mu\nu\alpha\beta} V_{\mu\nu} \delta_{\alpha\beta}^{(2)}(x_{\perp}).
\end{align}
where in the second line the anomalous current has been rewritten in a covariant form.
The zero mode non-conservation has exactly the same form as the bulk non-conservation equation, which can be seen by taking divergence of (\ref{current from action})

\begin{equation}\label{zm current div}
  \partial_{\mu}j_{bulk}^{\mu}= \frac{1}{8\pi^2} \epsilon^{\mu\nu\alpha\beta} V_{\mu\nu} \partial_{\alpha} \partial_{\beta}\varphi.
\end{equation}
Combining equations (\ref{zm current div}) and (\ref{4D current div}) leads to the conclusion that sum of bulk and zeromode currents are conserved.

\begin{equation}\label{current conserv}
  \partial_{\mu} (j_{bulk}^{\mu}+ j_{zm}^{\mu})=0.
\end{equation}
 Now we continue to show the implications of non-uniform rotation for quarks on vortex string interacting with superfluid velocity field.  The relation between Hamiltonian index and zero modes will be used to show that the number of zero modes change in a non-uniform rotation.
 Action (\ref{action vlocity coupl}) for the system with coupling to the Goldstone field $\phi$ becomes

\begin{equation}\label{action Goldston coupl}
  S=  \int d^4x ~\bar{\psi} (\partial_{\mu}+ i\partial_{\mu}\phi)\gamma^{\mu} \psi.
\end{equation}
So to incorporate rotation in the effective formalism we perform the identifications $\partial_0\phi \rightarrow \mu , \partial_i\phi \rightarrow \mu v_i^s$ with $v_i^s=(-\Omega y, \Omega x,0)$.  Nearby the vortex for the Goldstone field we have

\begin{equation}\label{vortex no vorticity}
  n= \frac{1}{2\pi} \int dx_i \partial_i \phi =  \frac{\mu}{2\pi} \int dx_i v_i^s = \frac{\mu}{2\pi} \int d^2x ~\Omega^3,
\end{equation}
 $n$ is the quantum number of vorticity and $\boldsymbol{\nabla} \times \textbf{v}^s=2 \boldsymbol{\Omega}$.
 We continue by relating the vortex quantum number to index of Hamiltonian. In this way, the angular momentum changes is related to the changes of net chiral fermions on the vortex ring. Following exactly the same steps of  \cite{Kirilin:2012mw, Metlitski:2005pr}, where using transverse Hamiltonian and index theorem, results in

\begin{equation}\label{index to n}
    N= \frac{1}{2\pi} \int dx_i \partial_i\phi =n.
 \end{equation}
 Now, equations (\ref{vortex no vorticity})and (\ref{index to n}) imply that the vortex quantum number is related to the index of transverse Hamiltonian. As pointed out at the beginning of this section, for the case of time-dependent rotation the Hamiltonian is not a constant of motion, but we can assume that an accelerating rotation can be done in sequential stages from one constant angular momentum to another infinitesimally close one.  Equation (\ref{vortex no vorticity})and (\ref{index to n}) give

\begin{equation}\label{index to derv n}
  \frac{dN}{dt}= \frac{\mu}{2\pi} \int d^2x \frac{d\Omega^3}{dt}.
\end{equation}
 It can be seen from (\ref{index to derv n}) that non-uniform rotation changes the number of chiral fermions on the vortex.
 So an increase or decrease of angular momentum leads to an increase or decrease of chiral fermions on the vortex ring. The reason is that as the massless fermions on the vortex ring move with the speed of light, to effectively change the angular momentum of vortex ring one has to change the number of fermions on the ring.

   The above mentioned mechanism can be used to give a microscopic explanation for cross-correlation between rotation and magnetic field.
 Although peripheral force field $F_{\theta}$ on vortex ring is charge blind, as the chiral fermions are charged, their  current  can generate a magnetic field. This is the microscopic explanation for cross-correlated response between rotation and magnetic field discussed in \cite{Huang:2017pqe}. Note that current non-conservation equation  (\ref{covar anom curr}) due to non-uniform rotation $\partial_{\alpha}j_{zm}^{\alpha} =  -(V_{01}/2\pi)  \delta^{(2)}(x_{\perp})= -(\mu \partial_t v_{\theta}/2\pi)  \delta^{(2)}(x_{\perp})$, is equivalent to (1+1)-dimensional electromagnetic gauge anomaly $\partial_{\alpha}j_{zm}^{\alpha} =  -(E_{z}/2\pi)  \delta^{(2)}(x_{\perp})$, which causes the charge to flow towards or from the bulk in both cases.

 \section{Rotating Frame}\label{rotat frame}

 Previous sections used the effective gauge field approach to study dense matter under rotation.
 In this section dense quark matter will be studied using gravitoelectromagnetic formalism and coordinate transformation to rotating frame. It confirms the previously obtained results from another point of view. Using the same approach a layered domain-wall structure was predicted in \cite{Huang:2017pqe} with surface angular momentum and baryon number densities. We will apply the formalism to the system of domain-wall and vortex ring. The gravitoelectromagnetic formalism shows that the previously obtained radial flow is a Hall type current.
 Cylindrical geometry is considered $(r, \theta, z)$ and the rotating metric tensor is of the form

\begin{equation}\label{rotation metric}
  ds^2=(1-r^2\Omega^2)dt^2 + r^2 d\theta^2 + r^2\Omega  d\theta dt + dz^2.
\end{equation}
 It can be seen from (\ref{rotation metric}) that coordinates transverse to the vortex ring $(r,z)$, are not affected by rotation.
 The anomalous term for two-flavor QCD at finite isospin chemical potential for neutral pion under rotation is

\begin{equation}\label{anomaly rotation pion}
  \mathcal{L}_{anom}= \frac{ \mu_I^2}{2\pi^2 f_{\pi}} \boldsymbol{\nabla}\varphi .\mathbf{\Omega},
\end{equation}
 where the above term accounts for the angular momentum density  and isospin number densities on the domain wall with nonzero $\boldsymbol{\nabla}\pi_0$

\begin{equation}\label{angular isosp density}
  \mathbf{j}=\frac{\mu_I^2}{2\pi^2 f_{\pi}} \langle\boldsymbol{\nabla}\varphi\rangle, ~~~~~ n_I= \frac{ \mu_I}{2\pi^2 f_{\pi}} \langle\boldsymbol{\nabla}\varphi \rangle .\mathbf{\Omega},
\end{equation}
 Now we discuss the radial current flow due to non-uniform rotation. There is a peripheral  anomalous Hall energy current discussed in \cite{Huang:2017pqe, Zubkov:2018gmc},  resulting from inhomogeneous chemical potential which is also present here. Our main focus is to discuss the relation between nonuniform rotation and radial energy flow using gravitoelectromagnetic formalism. The space-time metric has the gravitoelectromagnetic form

 \begin{equation}\label{GravEM metric}
   dS^2= c^2 (-1+ \frac{2\phi_g}{c^2})dt^2 - \frac{4}{c} \mathbf{A}_g.d\mathbf{x}dt + (1+ \frac{2\phi_g}{c^2})d\mathbf{x}.d\mathbf{x}.
 \end{equation}
 Where $\phi_g$ and $\textbf{A}_g$, in analogy with electromagnetism, are the scalar and vector potentials, i.e. gravitoelectromagnetic 4-potential $A_g^{\mu} =(\phi_g, \textbf{A}_g)$. The gravitoelectromagnetic fields are defined

 \begin{equation}\label{gravEM field}
   \mathbf{E_g}= -\boldsymbol{\nabla} \phi_g -\frac{1}{2c}\frac{\partial}{\partial t}(\mathbf{A_g}) ,   ~~~~~~  \mathbf{B_g}=\boldsymbol{\nabla} \times \mathbf{A_g}.
 \end{equation}
 The first gradient term of electric field equation  $\frac{\partial \mu_I(r)}{\partial r} \frac{\mathbf{r}}{r}$ is responsible for the anomalous Hall energy current  \cite{Huang:2017pqe, Zubkov:2018gmc}. We are  interested here in the second term  $\partial \mathbf{A_g}/\partial t$ which gives a radial Hall energy current. This peripheral gravitoelectric field results from accelerating or decelerating rotational motion. In the electromagnetic case a growing magnetic field generates a peripheral electric field via Faraday's law. Here a growing angular velocity $\Omega(t)$,  as can be seen from Eq. (\ref{rotation metric}), results in $\partial \mathbf{A_g}/\partial t= r\dot{\Omega}(t)$, which is time derivative of peripheral velocity . The drift velocity for Hall effect is orthogonal to both electric and magnetic field, i.e.  $\mathbf{j}_{Hall}\propto \mathbf{E}_g \times \mathbf{B}_g$. Then from $A_g=r\Omega=v_{\theta}$, Eq.'s (\ref{angular isosp density}),  (\ref{gravEM field}) and $\partial_z \varphi= \frac{\mu_I^2}{4 \pi^4 f_{\pi}^2}\Omega$ \cite{Huang:2017pqe}, so

 \begin{align}\label{E and B}
   \mathbf{E}_g= -\frac{1}{2c}\frac{\partial}{\partial t}(\mathbf{A_g}) = -\frac{1}{2c} \partial_t(A_g \hat{\boldsymbol{\theta}})=  -\frac{1}{2c} \partial_t v_{\theta}~ \hat{\boldsymbol{\theta}}     \nonumber     \\
   \mathbf{B}_g=\boldsymbol{\nabla} \times (A_g \hat{\boldsymbol{\theta}})= - \partial_r (r \Omega) ~ \hat{\mathbf{z}}= \frac{4 \pi^4 f_{\pi}^2}{\mu_I^2} ~ \partial_z \varphi ~ \hat{\mathbf{z}}
 \end{align}
 And a radial current is obtained as

 \begin{equation}\label{GEM radial current}
   \mathbf{j}_{Hall} ~\propto~ \mathbf{E}_g \times \mathbf{B}_g ~\propto ~  \partial_t v_{\theta}~ \partial_z \varphi ~ \hat{\boldsymbol{\theta}} \times \hat{\mathbf{z}}.
 \end{equation}
  For exact equality of the above relations one needs to introduce other parameters such as density of particle in moving frame, which we ignore for simplicity. Equation (\ref{GEM radial current}) is structurally the same as the term obtained from effective approach (\ref{radial current}).
  So in gravitoelectromagnetic formalism the radial flow is Hall type energy current.
  Here we add some comments on the gradient term of electric field in (\ref{gravEM field}), which gives rise to peripheral anomalous Hall current for inhomogeneous chemical potential under rotation. For an unlimited domain-wall with homogeneous chemical potential there would be no Hall energy current. But for the case of a limited  domain-wall, we argue that always a peripheral Hall current exists at the edge.
  This is because a nonzero angular velocity induces a charge density on the disc shaped domain wall. So even for homogeneous chemical potential under rotation, a sharp change would occur at the edge. This sharp change yields a radial chemical potential gradient which according to \cite{Huang:2017pqe, Zubkov:2018gmc}, results in a peripheral current at the domain-wall edge .

\section{Gravitational Anomaly Inflow}\label{grav anom}

 In this section we discuss gravitational anomaly and gravitational thermal Hall effect on domain wall resulting from a non-uniform rotation. The phenomena has no counterpart in the case with a background magnetic field.
 The anomaly results from non-invariance of the fermion determinant under general coordinate transformation \cite{AlvarezGaume:1983ig, Callan:1984sa }. First we give a microscopic explanation based on one-loop effective action of Weyl fermion propagating in a gravitational field. Then as a macroscopic manifestations of gravitational anomaly, we will consider an effective action for $(1+1)$-dimensional fermion zeromode coupled to background gravitational field .
   In fact in this section we will show that a non-uniform rotation guarantees the existence of pure gravitational anomaly so that the general formalisms can be applied.

  Microscopically, at one-loop level the analogue of gauge anomaly triangle diagram is the fermion energy-momentum two-point function \cite{AlvarezGaume:1983ig}. Anomaly arises because the effective action for metric perturbation coupled to the fermion energy-momentum tensor two-point function is not invariant under general coordinate transformation.
  The propagation of fermions in a weak gravitational field will be considered and the metric can be written in the form $g_{\mu\nu}=\eta_{\mu\nu} +h_{\mu\nu}$, where $\eta_{\mu\nu}$ is the Minkowski metric. At the linearized level, the interaction Lagrangian is of the form
  $\Delta \mathcal{L}=h^{\mu\nu}T_{\mu\nu}$, where $T_{\mu\nu}=\frac{1}{2}i \bar{\psi}(\gamma_{\mu} \overset\leftrightarrow{\partial}_{\nu}+\gamma_{\nu} \overset\leftrightarrow{\partial}_{\mu})\psi$ is the fermion energy-momentum tensor.
  Under infinitesimal general coordinate transformation $\delta x^{\mu}= \varepsilon^{\mu}$ the metric $h_{\mu\nu}$ transforms as $\delta h_{\mu\nu}=-\partial_{\mu}\varepsilon_{\nu}-\partial_{\nu}\varepsilon_{\mu}$.  A space-time dependent parameter along the vortex line can be generated through a non-uniform rotation which makes the transformation parameter time dependent. So an anomalous interacting Lagrangian $(\Delta \mathcal{L})$ is generated which violates the fermion energy-momentum conservation.
   Note that for a uniform rotation, no local space-time dependent coordinate transformation would be generated, and therefore no anomaly arises in the coupling between metric perturbation and fermion vortex.

  Now we continue to an effective description of the gravitational anomaly.
  As discussed in \cite{Stone:2012ud} a surface energy-momentum flow to the boundary edge Hall current cannot occur in the presence of a uniform bulk gravitational field.  The energy-momentum flux is proportional to the gradient of Ricci tensor. We will show that Ricci tensor for $(1+1)$-dimensional boundary with non-zero gradient exists for an accelerated rotation. The boundary gravitational anomaly and the domain-wall surface energy-momentum flux are related via the Callan-Harvey inflow mechanism i.e. a gravitational "Hall effect".
  The proportionality of CVE to the Chern-Simons current was discussed in
  \cite{Flachi:2017vlp, Stone:2018zel}. It was argued in  \cite{Stone:2018zel} that the gravitational Chern-Simons current has two components and it is not possible to simply identify it with CVE. In our case, however, only the radial flow is concerned which is perpendicular to axis of rotation, so we can use the equality of the current with Chern-Simons term.
 Note that for a uniformly rotating quark matter there is no energy-momentum flow due to gravitational anomaly where the gradient of Ricci tensor vanishes. The Ricci scalar for a rotating frame is \cite{Rizzi:Book}

\begin{equation}\label{Ricci rotation}
  R= \frac{-6 \Omega^2}{(1-r^2 \Omega^2)^2}
\end{equation}
 A nonvanishing  gradient of Ricci scalar of $(1+1)$-dimensional boundary theory is possible only for a time dependent angular velocity $\Omega(t)$.
 The gravitational anomaly is the failure of the covariant conservation law for energy-momentum tensor and for a general $(1+1)$-dimensional theory is of the following form \cite{Stone:2012ud}

\begin{equation}\label{1+1energy momen noncons}
  \nabla_{\mu}T^{\mu\nu} =  \frac{1}{96\pi ~\sqrt{|g|}} \epsilon^{\nu\sigma} \partial_{\sigma}R
\end{equation}
 where the only nonzero derivative of Ricci scalar is

\begin{equation}\label{Ricci derivative}
  \partial_t R = \frac{12\left(\Omega(t)+ r^2\Omega(t)^3\right)}{\left(1-r^2\Omega(t)^2 \right)^3}\dot{\Omega}(t)
\end{equation}
 The above non-conservation of energy-momentum tensor on the two-dimensional boundary vortex is linked to the inflow from higher dimension. The gravitational Chern-Simons functional leads to a radial energy-momentum flow to the boundary. The gravitational (2+1)-dimensional Chern-Simons term can be written in terms of Christoffel-symbol form $\Gamma_{\nu}^{\mu}=\Gamma_{\nu}^{\mu\rho}dx^{\rho}$ as

 \begin{equation}\label{CS christoffel}
   C[\Gamma]= \frac{1}{96\pi}\int_M  Tr \left\{\Gamma d\Gamma + \frac{2}{3} \Gamma^3 \right\},
 \end{equation}
 Variation of Chern-Simons functional due to a change in $\Gamma$

 \begin{equation}\label{gamma variation}
  \delta \Gamma_{\nu}^{\mu\rho}= \frac{1}{2} g^{\mu\lambda} (\nabla_{\nu}\delta g_{\lambda\rho}+ \nabla_{\rho}\delta g_{\nu\lambda} - \nabla_{\lambda}\delta g_{\nu\rho} )
 \end{equation}
then making use of the Riemann tensor properties unique to three dimensions, we have

\begin{equation}\label{dlata gamma}
  \delta C[\Gamma]= \frac{1}{48\pi} \int_M d^3x \sqrt{g} C^{\mu\nu} \delta g_{\mu\nu} + boundary ~ terms,
\end{equation}
where $C^{\mu\nu}$ is the Cotton tensor

\begin{equation}\label{cotton tensor}
  C^{\mu\nu}= - \frac{1}{2 \sqrt{g}} \left( \epsilon^{\rho\sigma\mu}\nabla_{\rho}R_{\sigma}^{\nu} + \epsilon^{\rho\sigma\nu}\nabla_{\rho}R_{\sigma}^{\mu}  \right)
\end{equation}
Then the energy-momentum tensor reads to be

\begin{equation}\label{energy cotton}
  T^{\mu\nu}= -\frac{1}{24\pi} C^{\mu\nu}.
\end{equation}
 Following \cite{Stone:2012ud}, the metric tensor of the rotating frame on the domain wall disc can be written in the form

\begin{equation}\label{disc metric}
  ds^2= dr^2 + g_{ab}(t,R)dx^a dx^b
\end{equation}
 the boundary is at $r=R$. Then the Ricci tensor coincides with the two-dimensional boundary Ricci tensor, and can be written

\begin{equation}\label{Ricci Rab}
  R_b^a = \delta_b^a R(t),   ~~~~   a,b=t, \theta
\end{equation}
 The time derivative of  Ricci tensor in non-uniform rotation $\Omega(t)$ gives the following flux of $\theta$-component energy-momentum tensor flow into the boundary vortex ring  at $R$

\begin{equation}\label{energy flow to ring}
  T^{r\theta}=\frac{1}{96\pi ~\sqrt{g}}  \epsilon^{r\theta t} \partial_t R
\end{equation}
 This is the radial energy-momentum flow on the (2+1)-dimensional bulk.  The energy-momentum tensor non-conservation of Eq. (\ref{1+1energy momen noncons}) is the (1+1)-dimensional boundary gravitational anomaly. Comparing (\ref{1+1energy momen noncons}) and (\ref{energy flow to ring}), it can be seen that the bulk radial inflow of energy-momentum tensor precisely accounts for the (1+1)-dimensional boundary ring gravitational anomaly. So there would be a radial energy-momentum inflow or outflow in the period of non-uniform rotation and the flow stops for uniform rotation. As explained in the introduction the gravitational anomaly discussed here is responsible for the term equivalent to the temperature dependent part of CVE for free fermions.

\section{Conclusions}

 We considered pion superfluid under rotation, with superfluidity induced by isospin chemical potential. It is known that there exists stable domain-walls in rotating pionic medium \cite{Son:2007ny}. It was pointed out that the phenomena occurring on domain-wall can be equivalently understood in a different way by considering the axial domain wall to be bounded by an axial vortex, where anomaly induced charges on the domain-wall can be related to the charges of vortex ring \cite{Son:2004tq}. The vortex ring surrounding the finite size domain-wall is a superfluid ring and is used to  study anomaly induced phenomena in the system. Hydrodynamic approximation is used to describe rotating superfluid and its topological defects such as vortices. The previous related works focused on the uniform rotation of the system and the prominent phenomena was induced charges (isospin, baryon, angular momentum) on the domain-wall \cite{Son:2004tq, Son:2007ny, Fukushima:2018ohd, Huang:2017pqe}. Also the induced currents in previous studies were peripheral in direction\cite{Huang:2017pqe}.

 In this work, for the first time we have considered the non-uniform rotation of the domain-wall and vortex system,  which leads to radial inflow or outflow of currents towards the boundary. We argued that the radial flow consists of gauge and gravitational anomaly related parts. The method used to obtain the triangle anomaly of effective theory relies on treating the local velocity field as an effective gauge theory. We showed that the domain-wall charges can not be conserved in presence of the vortex ring and the conserved quantity is the sum of domain-wall and vortex charges. Our main observation was that the radial current flows are prominent phenomena for non-uniform rotation,  where induced charges are the prominent phenomena for uniform rotation. Microscopic arguments for the gauge anomaly part and generation of fermionic modes is based on the effective Lagrangian and triangle anomaly with the generated electric field along the axis of rotation. The generated electric field is responsible for the fermion number non-conservation on the boundary which should be compensated from the bulk.  We discussed the agreement of the obtained results with the gravitoelectromagnetic formalism and coordinate transformation to a rotating frame metric. Also using the gravitoelectromagnetic formalism it was shown that the nature of radial flow due to gauge anomaly is a Hall type current. We pointed out that the peripheral current of \cite{Huang:2017pqe} is also present here due to the finite size effect of domain-wall at the edge. Another interesting new result was the existence of a pure gravitational anomaly on the $(1+1)$-dimensional theory of boundary vortex ring due to the non-uniform rotation. As a manifestation of Callan-Harvey anomaly inflow, the gravitational Chern-Simons term on domain-wall gives the required radial energy-momentum flow towards the boundary. Our microscopic calculation for the case of pure gravitational anomaly,  also confirms the effective theory approach. The microscopic picture of gravitational anomaly is based on fermion energy-momentum two point function and non-invariance of effective action under general coordinate transformation.  We mention that the radial flow on the domain-wall and vortex system discussed here, is perpendicular to the axis of rotation and consists of the gauge and pure gravitational anomaly related parts.



 \end{document}